\documentclass[aps,prd,showkeys,showpacs,amssymb,cite,
amsfonts,epsf,preprintnumbers,nofootinbib,superscriptaddress]{revtex4}

\usepackage[dvips]{graphicx}
\usepackage{bm,latexsym,amsmath,amssymb,amsfonts,color}

\newcommand{\be}{\begin{equation}}
\newcommand{\ee}{\end{equation}}
\newcommand{\bear}{\begin{eqnarray}}
\newcommand{\eear}{\end{eqnarray}}
\newcommand{\ba}{\begin{array}}
\newcommand{\ea}{\end{array}}

\begin{document}

\preprint{APCTP-Pre2015-013}

\title{Spectral indices in Eddington-inspired Born-Infeld inflation}

\author{Inyong Cho}
\email{iycho@seoultech.ac.kr}
\affiliation{Astroparticule et Cosmologie, Universit\'e Paris Diderot, 75013 Paris, France}
\affiliation{Institute of Convergence Fundamental Studies \& School of Liberal Arts,
Seoul National University of Science and Technology, Seoul 139-743, Korea}
\author{Jinn-Ouk Gong}
\email{jinn-ouk.gong@apctp.org}
\affiliation{Asia Pacific Center for Theoretical Physics, Pohang 790-784, Korea}
\affiliation{Department of Physics, Postech, Pohang 790-784, Korea}

\begin{abstract}

We investigate the scalar and tensor spectral indices of the quadratic
inflation model in Eddington-inspired Born-Infeld (EiBI) gravity.
We find that the EiBI corrections to the spectral indices are of second and first order
in the slow-roll approximation for the scalar and tensor perturbations respectively.
This is very promising since the quadratic inflation model in general relativity
provides a very nice fit for the spectral indices.
Together with the suppression of the tensor-to-scalar ratio
EiBI inflation agrees well with the observational data.

\end{abstract}

\maketitle

%%%%%%%%%%%%%%%%%%%%%%%%%%%%%%%%%%%%%%%%%%%%%%%%%%%%%%%%%%%%%%%%%%%%%%%%
\section{Introduction}

Inflation~\cite{inflation,Linde:1983gd} is considered
as the leading candidate to explain the otherwise extremely
finely tuned initial conditions in the very early Universe,
such as the horizon and flatness problems~\cite{Kolb:1990vq}.
Furthermore, during inflation quantum fluctuations are stretched to
super-horizon scales and frozen, which can naturally provide the origin of the
temperature fluctuations in the cosmic microwave background (CMB) on large scales~\cite{books}.
It is, however, a highly nontrivial task to realize inflation in a concrete
model based on high energy physics~\cite{Lyth:1998xn}.
One of the difficulties is that, in general, inflation is a highly sensitive probe
of higher dimensional operators (represented by for example
the eta problem~\cite{Copeland:1994vg})
and thus it requires an ultraviolet completion of the
effective theory~\cite{eft} in which inflation is described.
That is, we need quantum gravity
to accommodate inflation concretely. While the theory of quantum gravity is still
elusive, an alternative is to describe inflation in a theory of gravity which does not
require quantum aspects. An interesting candidate of this kind for theory is the
so-called Eddington-inspired Born-Infeld (EiBI) gravity~\cite{Banados:2010ix}.

An inflation model in the EiBI theory of gravity, which naturally avoids addressing the
quantum nature of gravity by construction, was developed only recently~\cite{Cho:2013pea}.
The model is based on a scalar field with a quadratic potential similar to the chaotic inflation
model~\cite{Linde:1983gd} in general relativity (GR). The primordial perturbations
of the model were investigated in Refs.~\cite{Cho:2014ija,Cho:2014jta,Cho:2014xaa,Cho:2015yza}.
In these works, the power spectra of both the scalar and tensor perturbations were studied,
and it was found that the tensor power spectrum can be suppressed significantly, while
the scalar spectrum remains almost the same. Therefore, the tensor-to-scalar ratio $r$ can
be suppressed significantly. In particular in the strong EiBI-gravity regime, $r$ can be even
further lowered close to zero. This fits the current observational constraints
$r_{0.05} < 0.12$~\cite{Ade:2015tva} and $r_{0.002} < 0.11$~\cite{Ade:2015lrj} at the $95\%$
confidence level.

This is very encouraging, because in the chaotic inflation model in GR with a power-law
potential, typically the tensor-to-scalar ratio is as large as $r = {\cal O}(0.1)$ so that
the quartic potential is almost ruled out and even the quadratic one is moderately
disfavored~\cite{Ade:2015lrj}. We should note, however, that although the chaotic
inflation model predicts too large a tensor-to-scalar ratio, it fits the spectral index very well.
The spectral index of the scalar power spectrum is very well constrained as
$n_{\cal R} = 0.968 \pm 0.006$ at the $68\%$ confidence level~\cite{Ade:2015lrj}, and is
another key parameter to judge the viability of a given model. For EiBI inflation to remain
as an attractive and viable alternative to the conventional chaotic inflation, the spectral
indices of the power spectra should be consistent with the current observational constraints.
In this article, we investigate the scalar and tensor spectral indices in EiBI inflation.

This article is organized as follows.
In Sec.~2, we present a summary of the inflationary feature of the quadratic model
in the EiBI theory of gravity investigated in Ref.~\cite{Cho:2013pea},
and the primordial perturbations in this model investigated
in Refs.~\cite{Cho:2014ija,Cho:2014jta,Cho:2014xaa,Cho:2015yza}.
In Sec.~3, we investigate the spectral indices of the model.
In Sec~4, we conclude.

\section{Summary of EiBI Inflation and Primordial Perturbations}

In this section, we summarize the inflationary feature investigated
in Ref.~\cite{Cho:2013pea},
and the scalar and tensor perturbations investigated
in Refs.~\cite{Cho:2014ija,Cho:2014jta,Cho:2014xaa,Cho:2015yza}.

\subsection{Inflation in EiBI Gravity}

The EiBI theory of gravity is described by the action~\cite{Banados:2010ix}
\begin{eqnarray}\label{action}
S_{{\rm EiBI}}=\frac{1}{\kappa}\int
d^4x\Big[~\sqrt{-|g_{\mu\nu}+\kappa
R_{\mu\nu}(\Gamma)|}-\lambda\sqrt{-|g_{\mu\nu}|}~\Big]+S_{\rm M}(g,\varphi),
\end{eqnarray}
where the matter action for inflation~\cite{Cho:2013pea} is given by
\be \label{S:chaotic}
S_{\rm M}(g,\varphi) = \int d^4 x \sqrt{-|g_{\mu\nu}|}
\left[ -\frac12 g_{\mu\nu} \partial^\mu\varphi \partial^\nu \varphi -V(\varphi) \right]
\quad
\text{with} \quad
V(\varphi) = \frac{m^2}{2} \varphi^2.
\ee
The gravity action is of Born-Infeld type, but it uses the Palatini formalism;
the metric $g_{\mu\nu}$ and the connection
$\Gamma_{\mu\nu}^{\rho}$ are regarded as independent fields.
The matter action is coupled to $g_{\mu\nu}$ only.
We set $8\pi G=1$, and $\kappa$ is the only additional parameter of EiBI theory.
The cosmological constant is related to the dimensionless parameter $\lambda$
by $\Lambda = (\lambda -1)/\kappa$,
and we consider the case of no cosmological constant ($\lambda=1$) in this article.

Performing a variation of the action \eqref{action}
with respect to the metric and the connection,
one can cast the equations of motion as
\begin{align}
\frac{\sqrt{-|q|}}{\sqrt{-|g|}}~q^{\mu\nu}
& =\lambda g^{\mu\nu} -\kappa T^{\mu\nu},\label{eom1}\\
q_{\mu\nu} & = g_{\mu\nu}+\kappa R_{\mu\nu}, \label{eom2}
\end{align}
where $q_{\mu\nu}$ is regarded as an auxiliary metric which provides
\be\label{Gamma}
\Gamma_{\alpha\beta}^{\mu}=\frac{1}{2}
q^{\mu\sigma}(q_{\alpha\sigma,\beta} + q_{\beta\sigma,\alpha} + q_{\alpha\beta,\sigma}).
\ee
The energy-momentum tensor is in the standard form,
$T^{\mu\nu} =(2/\sqrt{-|g|}) \delta L_{\rm M}/\delta g_{\mu\nu}$.
We take the metric ansatz as
\begin{align}
g_{\mu\nu}dx^\mu dx^\nu
= -dt^2 +a^2(t) \delta_{ij}dx^idx^j
= a^2(\eta) \left( -d\eta^2 +\delta_{ij}dx^idx^j \right),
\end{align}
where $t$ is the cosmological time and $\eta$ is the conformal time~\cite{Time}.
The scalar-field equation is then given by
\be\label{Seq}
\hat{\hat{\varphi}}_0+3H\hat{\varphi}_0 + \frac{dV}{d\varphi_0}
= \frac{1}{a^2} \left( \varphi_0''+2{\cal H}\varphi_0' + a^2\frac{dV}{d\varphi_0} \right) = 0 \, ,
\ee
where the hat and the prime denote a derivative with respect to $t$ and $\eta$, respectively,
and $H \equiv \hat{a}/a$ and ${\cal H} \equiv a'/a$.
The subscript $0$ stands for the unperturbed background field.

Due to the Born-Infeld (square-root) type of the gravity action \eqref{action},
there is an upper bound in pressure.
When the energy density is high,
the maximal pressure state (MPS) is achieved, $p_0 =\lambda/\kappa$,
beyond which the theory is not defined.
In the MPS, the Universe undergoes an exponential expansion.
The MPS is the past attractor
from which all the evolution paths of the Universe originate.
One may assume that the Universe approaches the MPS as $t\to -\infty$
in the past.
The energy density is high in the MPS, but
the gravitational curvature remains constant because $H_{\rm MPS} \approx 2m/3$.
This difference from GR comes from the difference of the Friedmann equation in EiBI gravity.
From the gravity point of view, therefore,
the quantum aspect is not necessary in describing the high-energy state
of the early Universe.

The MPS is unstable under a small perturbation,
and the Universe evolves to the so-called {\it near-MPS stage}
for which the background solutions were obtained in Ref.~\cite{Cho:2013pea}.
After the Universe leaves the near-MPS stage,
it enters into the {\it intermediate stage} followed by the {\it attractor stage}
in which the inflationary feature is very similar to that
of the chaotic inflation model in GR.
Therefore, there are two accelerating stages in EiBI inflation.

There are two ways in which the EiBI gravity effects are realized significantly.
First, when the matter density is high, so is the gravitational effect.
This corresponds to the near-MPS stage.
Second, if one makes the value of $\kappa$ arbitrarily large,
the gravity effect can appear strongly
even when the matter density is low.
At the attractor stage, the matter density is already low,
but EiBI gravity can be effective
if one turns on $\kappa$ strongly.
The following radiation- and matter-dominated epochs are
very similar to those in GR because the energy density is very low.

If the Universe spends a sufficient time at the attractor stage
and acquires 60 $e$-foldings, the EiBI prediction for CMB
will be very similar to that in GR, with only a very small correction.
However, if a sufficient number of $e$-foldings were not acquired at this stage,
the history at the near-MPS stage will be implied in the CMB at very long-wavelength scales.

In studying the primordial perturbations~\cite{Cho:2014jta,Cho:2014xaa,Cho:2014ija,Cho:2015yza},
therefore, the initial perturbations are considered to be produced
at the near-MPS stage.
They are assumed to evolve adiabatically until the attractor stage,
and to exit near the beginning of the attractor stage.
The coefficients of the mode functions are fixed
by imposing the initial conditions at the near-MPS stage.
The power spectra are evaluated at the horizon crossing at the attractor stage.
Here, we present the results of the scalar and the tensor perturbations briefly.

\subsection{Scalar Perturbation}

The scalar perturbation fields for the metrics are introduced as
\begin{align}
ds_q^2 &= b^2\left\{-\frac{1+ 2\phi_1}{z} d\eta^2  +2\frac{B_{1,i}}{\sqrt{z}} d\eta dx^i
+ \Big[(1-2\psi_1)\delta_{ij} + 2 E_{1,ij}\Big] dx^i dx^j \right\}, \label{qpert} \\
ds_g^2 &= a^2\left\{-(1+ 2\phi_2) d\eta^2 + 2 B_{2,i}d\eta dx^i
+\Big[(1-2\psi_2)\delta_{ij} + 2 E_{2,ij}\Big]dx^i dx^j \right\},  \label{metric_pert}
\end{align}
and the matter-field perturbation is
$\varphi=\varphi_0 +\chi$.
Here $a$, $b$, and $z$ are the background gravitational fields,
and from Eq.~\eqref{eom1} we get
\be\label{zb}
z=\frac{1+\kappa\rho_0}{1 - \kappa p_0}
\quad
\text{and} \quad
b = (1+\kappa\rho_0)^{1/4}(1-\kappa p_0)^{1/4}a,
\ee
where $\rho_0 = \varphi_0'^2/(2a^2) + V$
and $p_0 = \varphi_0'^2/(2a^2) - V$.

Imposing gauge conditions for the Fourier modes as $\psi_1=0$ and $E_1=0$,
all the perturbation fields for the metrics are expressed by
$\chi$ and the background fields.
In particular, the field $\psi_2$ that is used in evaluating the power spectrum
is given by
\be\label{psi2_XY}
\psi_2 = \frac{z-1}{2\kappa hz(z+1)(3z-1)}
\Big[ -2\kappa hz(z-1){\cal X}\chi' +a^2(z-1)^2{\cal X}\chi +2\kappa hz(3z-1){\cal Y}\chi \Big]
\approx \frac{z-1}{z+1} {\cal Y} \chi,
\ee
where $h\equiv b'/b$,
${\cal X} \equiv 1/(a\sqrt{\rho_0+p_0})$ and
${\cal Y} \equiv -m\sqrt{\rho_0-p_0}/(\rho_0+p_0)$,
and the third term in the square brackets is most dominant for the approximation
studied in Ref.~\cite{Cho:2015yza}.

The matter perturbation $\chi$ can be transformed to the canonical field $Q$ by
$Q \equiv \omega\chi$
together with the time transformation $d\tau \equiv (\omega^2/f_1)d\eta$,
where
\begin{align}\label{omega}
f_1 = \frac{3z^2-2z+3}{(z+1)(3z-1)}a^2
\quad
\text{and} \quad
\omega^4= \frac{3 z^2-2 z+3}{z(z+1)(3z-1)}a^4 \equiv W_0^4a^4.
\end{align}
Recalling that $a \approx 1/\left[\varphi_i m(\tau-\tau_0)\right]$ at the attractor stage,
the perturbation equation becomes
\be\label{Q-eq2}
\ddot{Q} + \left[ k^2 - \frac{2}{(\tau-\tau_0)^2} \right]Q
\approx 0,
\ee
where the dot denotes a derivative with respect to $\tau$,
$\varphi_i$ is the value of the scalar field at the beginning of the attractor stage,
and $\tau_0$ corresponds to the moment of the end of inflation.
At leading order, we have $d\tau \approx d\eta$
at the attractor stage.
This equation is the same as in GR,
and the solution is given by
\begin{align}
Q_{\rm ATT}(\tau)  &\approx A_1 \left\{ \cos \left[k(\tau-\tau_0)\right] - \frac{\sin \left[k(\tau-\tau_0)\right]}{k(\tau-\tau_0)} \right\}
+A_2 \left\{ \sin \left[k(\tau-\tau_0)\right] + \frac{\cos \left[k(\tau-\tau_0)\right]}{k(\tau-\tau_0)} \right\} \label{muATTmt1}
\\
&= \widetilde{A}_1 \left[1+ \frac{i}{k (\tau-\tau_0)}\right] e^{ik(\tau-\tau_0)}
+ \widetilde{A}_2  \left[1- \frac{i}{k (\tau-\tau_0)}\right] e^{-ik(\tau-\tau_0)},\label{muATTmt2}
\end{align}
where $\widetilde{A}_1 \equiv (A_1-i A_2)/2$ and $\widetilde{A}_2 \equiv (A_1+i A_2)/2$.
The coefficients $\widetilde{A}_i$ are to be determined by imposing the initial conditions
at the near-MPS stage, which makes it different from GR.
At the end of inflation, one gets an approximation,
$Q_{\rm ATT}(\tau) \approx i \left( \widetilde{A}_1-\widetilde{A}_2 \right)/[k(\tau-\tau_0)]$.

As studied in Refs.~\cite{Cho:2014jta,Cho:2014xaa,Cho:2015yza},
the minimum-energy condition is imposed at the initial moment $\tau_*$
of the perturbation production at the near-MPS stage~\cite{IP}.
Then the perturbations are assumed to evolve adiabatically
through the intermediate stage at which the WKB solution is applied,
and finally enter the attractor stage.
Performing the solution-matching at two transition moments
of three stages, one gets the coefficients $\widetilde{A}_i$
in terms of the near-MPS quantities expressed in $\tau_*$ as
\be
|Q_{\rm ATT}|^2 \approx
\frac{\left|\widetilde{A}_1-\widetilde{A}_2\right|^2}{k^2(\tau-\tau_0)^2}
\equiv \frac{D_k}{2k^3(\tau-\tau_0)^2}
\equiv D_k |Q_{\rm ATT}^{\rm GR}|^2,
\ee
where $|Q_{\rm ATT}^{\rm GR}|^2$ is the same as in GR,
and $D_k$ imprints the EiBI effect from the initial condition at the near-MPS stage,
\be\label{Dk}
D_k  \equiv 2k\left|\widetilde{A}_1-\widetilde{A}_2\right|^2
= \frac{2}{\pi} \left( c^2+R^2 + \frac{\pi^2}{16c^2} \right).
\ee
Here, $c$ and $R$ are determined from the initial condition as
\be\label{cR}
c^2 = \frac{\pi}4 \frac{Y^2 +Y_0^2}{|JY_0-J_0Y|}
\quad\text{and} \quad
R=\mp \sqrt{\frac{\pi}{4}}\frac{JY + J_0 Y_0}{\sqrt{|JY_0-J_0Y|(Y^2 +Y_0^2)}},
\ee
where $J \equiv {(J_0-2k\tau_* J_1)}/{\sqrt{1+ 4k^2\tau_*^2}}$,
$Y \equiv {(Y_0-2k\tau_* Y_1)}/{\sqrt{1+ 4k^2\tau_*^2}}$,
$J_{0,1} \equiv J_{0,1}(k\tau_*)$, and $Y_{0,1} \equiv Y_{0,1}(k\tau_*)$.

The comoving curvature perturbation is defined as
\be
{\cal R} = \psi_2 +\frac{H}{\hat{\varphi}_0}\chi,
\ee
and the scalar power spectrum evaluated at the horizon crossing becomes
\begin{align}
{\cal P}_{\cal R} &= \frac{k^3}{2\pi^2}{|\cal R|}^2
\approx \frac{D_k}{W_0^2}
\left| 1+ \frac{z-1}{z+1}\frac{\hat{\varphi}_0}{H}{\cal Y} \right|^2
\frac{k^3}{2\pi^2} \frac{H^2}{\hat\varphi_0^2} \frac{|Q^{\rm GR}_{\rm ATT}|^2}{a^2}
\equiv D_k \times E^{\rm S} \times {\cal P}_{\cal R}^{\rm GR}. \label{EkS}
\end{align}
Here, ${\cal P}_{\cal R}^{\rm GR} \equiv k^3H^2|Q^{\rm GR}_{\rm ATT}|^2/(2\pi^2\hat{\varphi}_0^2a^2)$
is the power spectrum in GR, and $E^{\rm S} \equiv
\left| 1+ (z-1)\hat{\varphi}_0{\cal Y}/[(z+1)H] \right|^2 /W_0^2$
is the EiBI correction.
The EiBI correction $E^{\rm S}$ applies for all scales,
while the correction $D_k$ manifests only for the long-wavelength modes.

\subsection{Tensor Perturbation}

The tensor perturbation fields $\gamma_{ij}$ and $h_{ij}$
are introduced as
\begin{align}
ds_q^2 &= -X^2 d\eta^2 +Y^2\left(\delta_{ij}+\gamma_{ij}\right)dx^i dx^j
= Y^2 \left[ -d\tau^2 +\left(\delta_{ij}+\gamma_{ij}\right)dx^i dx^j \right],\label{eqmunu} \\
ds_g^2 &= a^2 \left[ -d\eta^2 +\left(\delta_{ij} +h_{ij}\right) dx^i dx^j \right],\label{gmunu}
\end{align}
where $\tau$ is the conformal time for the auxiliary metric.
At the attractor stage, $\tau$ is the same as that for the scalar perturbation
in the leading order, $d\tau \approx d\eta$.
We impose the transverse and traceless conditions
on both $h_{ij}$ and $\gamma_{ij}$, i.e.,
${\partial}_{i}h^{ij}={\partial}_{i}\gamma^{ij}=0$ and $h^i{}_i=\gamma^i{}_i=0$.
From Eq.~\eqref{eom1}, one then gets $\gamma_{ij} = h_{ij}$ \cite{EscamillaRivera:2012vz},
and also
\be\label{Y}
X= \frac{(1-\kappa p_0)^{3/4}}{(1+\kappa\rho_0)^{1/4}}a
\quad
\text{and} \quad
Y=(1+\kappa\rho_0)^{1/4} (1-\kappa p_0)^{1/4}a \equiv Y_0a.
\ee
The Fourier mode for the perturbation is defined by
\be
h_{ij}(\eta,\vec{x}) =\sum_{\sigma = +,-}
\int \frac{d^3k}{(2\pi)^{3/2}} \;
h_{\sigma} (\eta,\vec{k}) \;
\epsilon^{\sigma}_{ij}(\vec{k}) \;
e^{i\vec k\cdot\vec x},
\ee
where $\epsilon^{\sigma}_{ij}$ is the polarization tensor.
Introducing a canonical field by $\mu_\sigma \equiv (Y/2)h_\sigma$,
from Eq.~\eqref{eom2} the perturbation equation is given by
\be\label{mueq}
\ddot\mu_\sigma + \left( k^2 -\frac{\ddot{Y}}{Y} \right)  \mu_\sigma = 0.
\ee
At the attractor stage, $d\tau \approx d\eta$
and $\ddot{Y}/Y \approx \ddot{a}/a$.
Therefore, Eq.~\eqref{mueq} is the same as in GR,
\be\label{mueqATT}
\ddot\mu_\sigma + \left[ k^2 -\frac{2}{(\tau-\tau_0)^2} \right] \mu_\sigma  \approx 0,
\ee
and the mode solution is
\be\label{muATT}
\mu_{\rm ATT}(\tau)  \approx A_1 \left\{ \cos \left[k(\tau-\tau_0)\right] - \frac{\sin \left[k(\tau-\tau_0)\right]}{k(\tau-\tau_0)} \right\}
+A_2 \left\{ \sin \left[k(\tau-\tau_0)\right] + \frac{\cos \left[k(\tau-\tau_0)\right]}{k(\tau-\tau_0)} \right\}.
\ee
Note that this solution has exactly the same form as Eq.~\eqref{muATTmt1}.
As investigated in Ref.~\cite{Cho:2014ija},
the near-MPS solution is also the same as that for the scalar perturbation.
Then, imposing the same initial conditions and
performing the solution matching in the same way,
we get
\be
|\mu_{\rm ATT}|^2 \approx D_k |\mu_{\rm ATT}^{\rm GR}|^2,
\ee
where $D_k$ is the same as in Eq.~\eqref{Dk},
and $|\mu_{\rm ATT}^{\rm GR}|^2$ is the value in GR.

The power spectrum at the horizon-crossing is
\begin{align}
{\cal P}_{\rm T}
&= \frac{k^3}{2\pi^2} |h_\sigma|^2
= \frac{D_k}{Y_0^2} \frac{2k^3}{\pi^2} \frac{|\mu_{\rm ATT}^{\rm GR}|^2}{a^2}
\equiv D_k \times E^{\rm T} \times {\cal P}_{\rm T}^{\rm GR}.\label{PT}
\end{align}
Here, ${\cal P}_{\rm T}^{\rm GR} \equiv 2k^3|\mu_{\rm ATT}^{\rm GR}|^2/\pi^2a^2$ is the spectrum in GR,
and $E^{\rm T} \equiv 1/Y_0^2$ is the EiBI correction applied for all scales.

\section{Spectral Indices}

In this section, we investigate the scalar and tensor spectral indices.
We consider two limits;
the weak and strong EiBI gravity limits, which correspond to
$\kappa \ll m^{-2}$ and $\kappa \gg m^{-2}$, respectively.

\subsection{Scalar Spectral Index}

The scalar spectral index is evaluated as
\be
n_{\cal R} - 1 \equiv \frac{d\log{\cal P}_{\cal R}}{d\log k}
= \frac{d\log{\cal P}_{\cal R}^{\rm GR}}{d\log k} + \frac{d\log E^{\rm S}}{d\log k}
+ \frac{d\log D_k}{d\log k} .
\ee
The second term was given earlier by
\be\label{ES}
E^{\rm S} = \frac{1}{W_0^2} \left| 1+ \frac{z-1}{z+1}\frac{\hat{\varphi}_0}{H}{\cal Y} \right|^2
\equiv (1+S_1) \left| 1-S_2 \right|^2,
\ee
where $S_1\equiv 1/W_0^2-1$ and
$S_2 \equiv -(z-1)\hat{\varphi}_0{\cal Y}/[(z+1)H] \approx -\hat{\varphi}_0\psi_2/(H\chi)$.

As studied in Ref.~\cite{Cho:2015yza},
at the attractor stage with the first slow-roll condition $\hat{\varphi}_0^2/2 \ll m^2\varphi_0^2/2$,
the first Friedmann equation in EiBI gravity is approximated by that in GR, $H^2 \approx  V/3$.
The scalar-field equation is the same,
so the second Friedmann equation is also approximated by that in GR.
Therefore, the slow-roll parameters in EiBI gravity are defined in the same way,
\be\label{epsilon12}
\epsilon_1 \equiv -\frac{\hat{H}}{H^2} \approx \frac{\hat{\varphi}_0^2}{2H^2}
\quad
\text{and} \quad
\epsilon_2 \equiv \frac{\hat{\epsilon}_1}{H\epsilon_1}.
\ee
At the attractor stage with the first slow-roll condition
we get from Eq.~\eqref{zb}
\begin{eqnarray}
z
= \frac{1+\kappa (\hat{\varphi}_0^2/2 + m^2\varphi_0^2/2)}{1-\kappa (\hat{\varphi}_0^2/2 - m^2\varphi_0^2/2)}
\approx 1 + \frac{\kappa \hat{\varphi}_0^2}{1+\kappa m^2\varphi_0^2/2}
\approx
\left\{
  \begin{array}{ll}
    1+ \kappa \hat{\varphi}_0^2 & \hbox{($\kappa \ll m^{-2}$)} \, , \\
    1+ \dfrac{2\hat{\varphi}_0^2}{m^2\varphi_0^2} & \hbox{($\kappa \gg m^{-2}$)} \, .
  \end{array}
\right.
\end{eqnarray}
Using this result, we get from Eq.~\eqref{omega}
\begin{eqnarray}
\omega^4
\approx
\left\{
  \begin{array}{lll}
    (1-2\kappa \hat{\varphi}_0^2)a^4
    & \approx (1-4\kappa H^2 \epsilon_1)a^4 & \hbox{($\kappa \ll m^{-2}$)} \, , \\
    \left( 1-\dfrac{4\hat{\varphi}_0^2}{m^2\varphi_0^2} \right)a^4 & \approx \left( 1-\dfrac{4}{3}\epsilon_1 \right)a^4 & \hbox{($\kappa \gg m^{-2}$)} \, ,
  \end{array}
\right.
\end{eqnarray}
which gives
\begin{eqnarray}
S_1
\approx
\left\{
  \begin{array}{ll}
    2\kappa H^2 \epsilon_1 & \hbox{($\kappa \ll m^{-2}$)} \, ,\\
    \dfrac{2}{3}\epsilon_1 & \hbox{($\kappa \gg m^{-2}$)} \, .
  \end{array}
\right.
\end{eqnarray}
Likewise,
at the attractor stage with the first slow-roll condition,
we get from Eq.~\eqref{psi2_XY} using $z$ in Eq.~\eqref{zb}
\be
\psi_2
\approx -\frac{\kappa m\sqrt{\rho_0-p_0}}{2+\kappa(\rho_0-p_0)} \chi
\approx -\frac{\sqrt{3}\kappa m H}{\sqrt{2}(1+3\kappa H^2)}\chi.
\ee
Then we get from the definition below Eq.~\eqref{ES}
\begin{eqnarray}
S_2 \approx \frac{\sqrt{3}\kappa m\hat{\varphi}_0}{\sqrt{2}(1+3\kappa H^2)} .
\end{eqnarray}
With the above results of $S_1$ and $S_2$,
we find $E^{\rm S}$ at the leading order as
\begin{equation}\label{ES}
E^{\rm S} \approx \left\{
\begin{array}{ll}
1 -\dfrac{\sqrt{6}\kappa m\hat{\varphi}_0}{1+3\kappa H^2}
+2\kappa H^2\epsilon_1 & \hbox{($\kappa \ll m^{-2}$)} \, ,
\\
1 -\dfrac{\sqrt{2}m\hat{\varphi}_0}{\sqrt{3}H^2}
+\dfrac{2}{3}\epsilon_1 & \hbox{($\kappa \gg m^{-2}$)} \, .
\end{array}
\right.
\end{equation}
Note that $m \approx \sqrt{3/2}\hat\varphi_0 \approx H\sqrt{3\epsilon_1}$
from the background scalar-field solution $\varphi_0 = \varphi_i +\sqrt{2/3}mt$
at the attractor stage.
Note also that $\kappa H^2 \approx \kappa m^2\varphi_0^2/2$
which can be as large as ${\cal O}(1)$ if $\kappa m^2 \sim 10^{-2}$.
Therefore, this is not entirely negligible in the limit of $\kappa \ll m^{-2}$.
We find the corresponding EiBI correction to the spectral index as
\begin{equation}\label{DES}
\frac{d\log E^{\rm S}}{d\log k} \approx \left\{
\begin{array}{ll}
-\kappa H^2 \left( 2-\dfrac{3}{1+3\kappa H^2} \right) \epsilon_1 ( 2\epsilon_1-\epsilon_2 )
-\dfrac{36\kappa^2 H^4}{(1+3\kappa H^2)^2} \epsilon_1^2
& \hbox{($\kappa \ll m^{-2}$)}
\quad \approx \kappa H^2\epsilon_1 ( 2\epsilon_1-\epsilon_2 )
\quad \hbox{($\kappa \ll H^{-2}$)} \, ,
\\
-\epsilon_1 \left( 2\epsilon_1 + \dfrac{5}{3}\epsilon_2 \right) & \hbox{($\kappa \gg m^{-2}$)} \, .
\end{array}
\right. 
\end{equation}
Here, we have used the relations at the horizon crossing,
$k=aH\approx -1/(\tau-\tau_0)$ and
$d/dk \approx (aH)^{-2}d/d\tau \approx a^{-1}H^{-2} d/dt$,
and $\hat{\hat{\varphi}}_0 \approx \hat{\varphi}_0 H (\epsilon_2/2-\epsilon_1)$
derived from the slow-roll parameters in Eq.~\eqref{epsilon12}.
Compared with the standard GR contribution,
\begin{equation}
\frac{d\log{\cal P}_{\cal R}^{\rm GR}}{d\log k}  = -2\epsilon_1-\epsilon_2,
\end{equation}
the EiBI correction in Eq.~\eqref{DES} is of second order in $\epsilon_i$,
and thus does not change the spectral index
of the scalar power spectrum at leading order.

\begin{figure*}[btph]
\begin{center}
 \includegraphics[width=.43\textwidth]{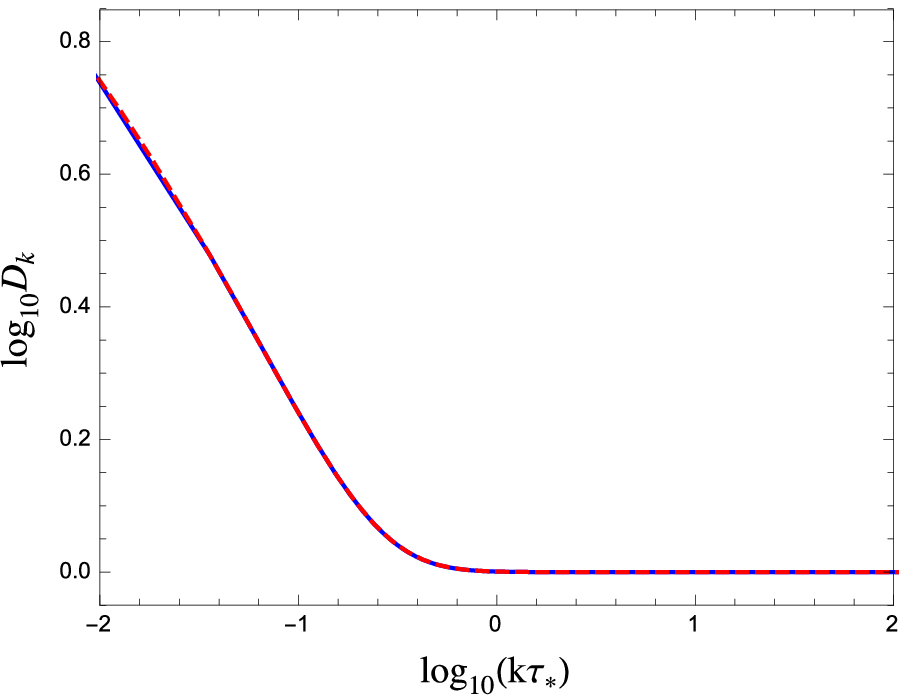}
 \hspace{1em}
 \includegraphics[width=.436\textwidth]{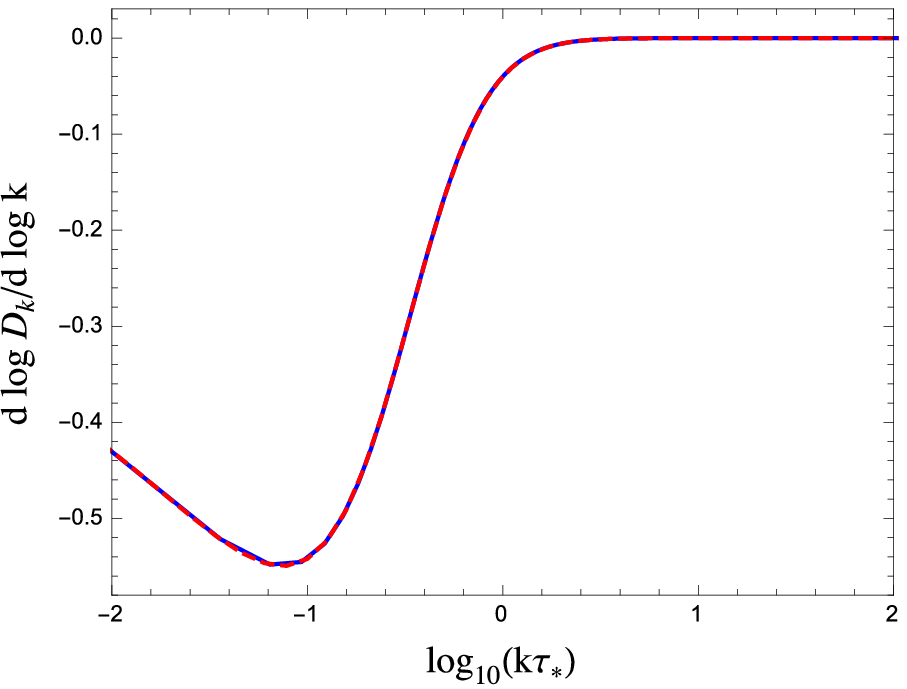}
\end{center}
\caption{Numerical plot of $D_k$ (left panel) and $d\log D_k/d\log k$ (right panel).
The parameters for the plots are
$a_0=1$, $m=10^{-5}$, and $\kappa=10^{13}$ (solid lines) and $\kappa = 10^9$ (dashed lines).
The value of $\psi_0$ was chosen from the relation in Ref.~\cite{IP}
to give $k\tau_*=1$ for $k=10^{-60}/l_p \approx 0.002/{\rm Mpc}$.}
\label{FIG1}
\end{figure*}

Another EiBI correction factor $D_k$ given by Eq.~\eqref{Dk} is explicitly $k$-dependent
and exhibits a peculiar rise at low $k$, while it approaches 1 in the high-$k$ region.
$D_k$ does not contribute to the tensor-to-scalar ratio
since it is common to both the scalar and tensor power spectra,
as can be seen from Eqs.~\eqref{EkS} and \eqref{PT}.
However, in principle it may significantly contribute to the spectral index.
This is especially worrisome if $\tau_*$, the initial moment of the perturbation production,
is well within the last 60 $e$-folds of the inflationary stage.
In Fig.~\ref{FIG1}, we show the behavior
of both $D_k$ and $d\log D_k/d\log k$ as functions of $k\tau_*$.
As can be seen, as long as we can push $\tau_*$ to satisfy $k\tau_* \gtrsim 1$
for the regime of our observational interest $k\gtrsim0.002/{\rm Mpc}$,
the contributions from $D_k$ to the power spectrum
and especially to the spectral index can be made negligible.
Note also that in Fig.~\ref{FIG1}
we show both the large- and small-$\kappa$ limits,
denoted by solid and dashed lines respectively.
The two cases are almost identical, so $D_k$ is not sensitive to the strength
of EiBI gravity while the other EiBI correction $E^{\rm S}$ is.
This is because the effect of $D_k$ originates from the near-MPS stage
at which the EiBI-gravity effect is strong due to the high matter-energy density.

%%%%%%%%%%%%%%%%%%%%%%%%%%%%%%%%%%%%%%%%%%%%%%%%%%%%%%%%%%%%%%
%%%%%%%%%%%%%%%%%%%%%%%%%%%%%%%%%%%%%%%%%%%%%%%%%%%%%%%%%%%%%%

\subsection{Tensor Spectral Index}

The tensor spectral index is evaluated as
\be
n_{\rm T} \equiv \frac{d\log{\cal P}_{\rm T}}{d\log k}
= \frac{d\log{\cal P}_{\rm T}^{\rm GR}}{d\log k} + \frac{d\log E^{\rm T}}{d\log k}
+ \frac{d\log D_k}{d\log k} .
\ee
At the attractor stage, we get from Eq.~\eqref{Y}
\be\label{ET}
E^{\rm T}
\approx \frac{1}{1+\kappa\rho_0}
\approx \frac{1}{1+3\kappa H^2}.
\ee
Following similar steps as for the scalar spectral index,
at the leading order in the slow-roll parameters we can easily find
the EiBI correction to the spectral index,
\begin{equation}
\frac{d\log E^{\rm T}}{d\log k} \approx \frac{6\kappa H^2}{1+3\kappa H^2}\epsilon_1 \approx
\left\{
\begin{array}{ll}
0 & \hbox{($\kappa \ll H^{-2}$)} \, ,
\\
2\epsilon_1 & \hbox{($\kappa \gg m^{-2}$)} \, ,
\end{array}
\right. 
\end{equation}
which is ${\cal O}(\epsilon)$.
This is very different from the scalar spectral index
for which we found that the EiBI correction
from $E^{\rm S}$ is ${\cal O}(\epsilon^2)$.
With the standard GR contribution,
\begin{equation}
\frac{d\log{\cal P}_{\rm T}^{\rm GR}}{d\log k} = -2\epsilon_1 \, ,
\end{equation}
the tensor spectral index is
\begin{equation}
n_{\rm T} \approx
\left\{
\begin{array}{ll}
-2 \left( 1- \dfrac{3\kappa H^2}{1+3\kappa H^2}  \right) \epsilon_1
& \hbox{($\kappa \ll m^{-2}$)}
\quad \approx -2\epsilon_1
\quad \hbox{($\kappa \ll H^{-2}$)} \, ,
\\
0 & \hbox{($\kappa \gg m^{-2}$)} \, ,
\end{array}
\right. 
\end{equation}
where the $D_k$ part is the same as for the scalar index.

\section{Conclusions}

In this article, we investigated in EiBI gravity the spectral indices of the the
primordial perturbations from the inflation model with a quadratic potential.
The result shows that the EiBI correction to GR is
of second order in the slow-roll parameters for the scalar spectral index.
Since the chaotic inflation model in GR with a power-law potential
provides a very good fit for the spectral index which is of first order,
this small EiBI correction is very affirmative in considering the viability of the model.
There is a running from the peculiar rise in the power spectrum for low-$k$ modes,
but it can be pushed to very large scales
with a proper choice of parameters.
The EiBI correction for the tensor spectral index is, however, of first order.
We have not yet detected the primordial tensor perturbations and thus our
EiBI-corrected tensor spectral index is still within the observational bound.
However, this is a verifiable prediction once $r$ is detected, which could be
accomplished in the next decade by future observations of the CMB
$B$-mode polarization. Indeed the sensitivity of the planned observations
is very ambitious; for example, LiteBIRD~\cite{Matsumura:2013aja} is
expected to give $r = {\cal O}(0.01-0.001)$.

The chaotic inflation model in GR predicts a quite large value of the tensor-to-scalar ratio,
so it is disfavored by observational results.
In EiBI inflation, however, the value can be reduced,
as investigated in Refs.~\cite{Cho:2014jta,Cho:2014xaa,Cho:2015yza}.
From Eqs.~\eqref{EkS} and \eqref{PT},
the tensor-to-scalar ratio is given by
\be
r \approx \frac{E^{\rm T}}{E^{\rm S}} r^{\rm GR},
\ee
where we can evaluate Eqs.~\eqref{ES} and \eqref{ET} further as
\begin{equation}
E^{\rm S} \approx \left\{
\begin{array}{ll}
1 -\dfrac{2\kappa m^2}{1+\kappa m^2\varphi_i^2/2}
+\dfrac{2}{3}\kappa m^2 & \hbox{($\kappa \ll m^{-2}$)}
\\
1 -\dfrac{8}{3\varphi_i^2}
& \hbox{($\kappa \gg m^{-2}$)}
\end{array}
\right.
\quad
\text{and} \quad
E^{\rm T} \approx \frac{1}{1+\kappa m^2\varphi_i^2/2} \, .
\end{equation}
In both limits of $\kappa$, the corrections for $E^{\rm S}$ are tiny
while $E^{\rm T}$ can be significantly lowered.
In the large-$\kappa$ limit particularly,
$E^{\rm T}$ can be reduced close to zero.
Therefore, the tensor-to-scalar ratio can be well within the observational bound.
Together with the results of this work on the spectral indices,
the quadratic inflation model in EiBI gravity is promising.

For the past several years, EiBI theory has been investigated in
cosmological and astrophysical aspects.
The density perturbations in the Friedmann universe driven by a perfect fluid were investigated
in Refs.~\cite{Lagos:2013aua,EscamillaRivera:2012vz,Avelino:2012ue,Yang:2013hsa}.
The observational bound for the theory parameter $\kappa$ was
obtained from the star-formation studies in Refs. \cite{Avelino:2012ge,Pani:2011mg,Pani:2012qb},
$|\kappa| < 10^{-2} \rm{m^5 kg^{-1}s^{-2}} \sim 10^{77}$ in Planck units.
A similar bound was also obtained from the study of atomic nuclei \cite{Avelino:2012qe}.
Both the weak- and strong-gravity results in our work are well within this bound,
as discussed in Refs. \cite{Cho:2014xaa,Cho:2015yza}.
One problem in EiBI gravity to be resolved is the surface singularity
accompanied with a polytropic star \cite{Pani:2012qd},
which arises in the similar pathology of Palatini $f(R)$ gravity theory \cite{Sotiriou:2008rp}.
One possible way of solving this might be to consider
the gravitational backreaction of the matter dynamics, as investigated in Ref. \cite{Kim:2013nna}.
This issue requires further investigation.

\section*{Acknowledgements}

This work was supported by grants from the National Research Foundation
funded by the Korean government, No. NRF-2012R1A1A2006136 (I.C.)
and No. NRF-2013R1A1A1006701 (J.G.).
J.G. is grateful to APC, Universit\'e Paris Diderot for hospitalities
while this work was in progress,
and acknowledges the Max-Planck-Gesellschaft, Gyeongsangbuk-Do and Pohang City
for the support of the Independent Junior Research Group
at the Asia Pacific Center for Theoretical Physics.


\begin{thebibliography}{99}



\bibitem{inflation}
%\cite{Starobinsky:1980te}
%\bibitem{Starobinsky:1980te}
  A.~A.~Starobinsky,
  %``A New Type of Isotropic Cosmological Models Without Singularity,''
  Phys.\ Lett.\ B {\bf 91}, 99 (1980)~;
  %%CITATION = PHLTA,B91,99;%%
%\cite{Guth:1980zm}
%\bibitem{Guth:1980zm}
  A.~H.~Guth,
  %``The Inflationary Universe: A Possible Solution to the Horizon and Flatness Problems,''
  Phys.\ Rev.\ D {\bf 23}, 347 (1981)~;
  %%CITATION = PHRVA,D23,347;%%
%\cite{Albrecht:1982wi}
%\bibitem{Albrecht:1982wi}
  A.~Albrecht and P.~J.~Steinhardt,
  %``Cosmology for Grand Unified Theories with Radiatively Induced Symmetry Breaking,''
  Phys.\ Rev.\ Lett.\  {\bf 48}, 1220 (1982).
  %%CITATION = PRLTA,48,1220;%%


%\cite{Linde:1983gd}
\bibitem{Linde:1983gd}
  A.~D.~Linde,
  %``Chaotic Inflation,''
  Phys.\ Lett.\ B {\bf 129} (1983) 177.
  %%CITATION = PHLTA,B129,177;%%


%\cite{Kolb:1990vq}
\bibitem{Kolb:1990vq}
  E.~W.~Kolb and M.~S.~Turner,
  %``The Early Universe,''
  Front.\ Phys.\  {\bf 69}, 1 (1990).
  %%CITATION = FRPHA,69,1;%%


\bibitem{books}
See e.g.
%\cite{Mukhanov:2005sc}
%\bibitem{Mukhanov:2005sc}
  V.~Mukhanov,
  %``Physical foundations of cosmology,''
  Cambridge, UK: Univ. Pr. (2005) 421 p~;
%\cite{Weinberg:2008zzc}
%\bibitem{Weinberg:2008zzc}
  S.~Weinberg,
  %``Cosmology,''
  Oxford, UK: Oxford Univ. Pr. (2008) 593 p.


%\cite{Lyth:1998xn}
\bibitem{Lyth:1998xn}
  D.~H.~Lyth and A.~Riotto,
  %``Particle physics models of inflation and the cosmological density perturbation,''
  Phys.\ Rept.\  {\bf 314}, 1 (1999)
  [hep-ph/9807278].
  %%CITATION = HEP-PH/9807278;%%


%\cite{Copeland:1994vg}
\bibitem{Copeland:1994vg}
  E.~J.~Copeland, A.~R.~Liddle, D.~H.~Lyth, E.~D.~Stewart and D.~Wands,
  %``False vacuum inflation with Einstein gravity,''
  Phys.\ Rev.\ D {\bf 49}, 6410 (1994)
  [astro-ph/9401011].
  %%CITATION = ASTRO-PH/9401011;%%


\bibitem{eft}
%\cite{Cheung:2007st}
%\bibitem{Cheung:2007st}
  C.~Cheung, P.~Creminelli, A.~L.~Fitzpatrick, J.~Kaplan and L.~Senatore,
  %``The Effective Field Theory of Inflation,''
  JHEP {\bf 0803}, 014 (2008)
  [arXiv:0709.0293 [hep-th]]~;
  %%CITATION = ARXIV:0709.0293;%%
%\cite{Weinberg:2008hq}
%\bibitem{Weinberg:2008hq}
  S.~Weinberg,
  %``Effective Field Theory for Inflation,''
  Phys.\ Rev.\ D {\bf 77}, 123541 (2008)
  [arXiv:0804.4291 [hep-th]].
  %%CITATION = ARXIV:0804.4291;%%
See for an alternative approach e.g.
%\cite{Achucarro:2012sm}
%\bibitem{Achucarro:2012sm}
  A.~Achucarro, J.~O.~Gong, S.~Hardeman, G.~A.~Palma and S.~P.~Patil,
  %``Effective theories of single field inflation when heavy fields matter,''
  JHEP {\bf 1205}, 066 (2012)
  [arXiv:1201.6342 [hep-th]].
  %%CITATION = ARXIV:1201.6342;%%


%\cite{Banados:2010ix}
\bibitem{Banados:2010ix}
  M.~Banados and P.~G.~Ferreira,
  %``Eddington's theory of gravity and its progeny,''
  Phys.\ Rev.\ Lett.\  {\bf 105}, 011101 (2010)
  [arXiv:1006.1769 [astro-ph.CO]].
  %%CITATION = ARXIV:1006.1769;%%


%\cite{Cho:2013pea}
\bibitem{Cho:2013pea}
  I.~Cho, H.~-C.~Kim and T.~Moon,
  %``Precursor of Inflation,''
  Phys.\  Rev.\  Lett 111, {\bf 071301} (2013)
  [arXiv:1305.2020 [gr-qc]].
  %%CITATION = ARXIV:1305.2020;%%


%\cite{Cho:2014ija}
\bibitem{Cho:2014ija}
  I.~Cho and H.~-C.~Kim,
  %``Inflationary Tensor Perturbation in Eddington-inspired Born-Infeld gravity,''
  Phys.\ Rev.\ D {\bf 90}, 024063 (2014)  [arXiv:1404.6081 [gr-qc]].
  %%CITATION = ARXIV:1404.6081;%%


%\cite{Cho:2014jta}
\bibitem{Cho:2014jta}
  I.~Cho and N.~K.~Singh,
  %``Tensor-to-Scalar Ratio in Eddington-inspired Born-Infeld Inflation,''
  Eur.\ Phys.\ J.\ C {\bf 74}, no. 11, 3155 (2014)  [arXiv:1408.2652 [gr-qc]].
  %%CITATION = ARXIV:1408.2652;%%


%\cite{Cho:2014xaa}
\bibitem{Cho:2014xaa}
  I.~Cho and N.~K.~Singh,
  %``Scalar Perturbation Produced at the Pre-inflationary Stage in Eddington-inspired Born-Infeld Gravity,''
  Eur.\ Phys.\ J.\ C {\bf 75}, no. 6, 240 (2015)  [arXiv:1412.6344 [gr-qc]].
  %%CITATION = ARXIV:1412.6344;%%


%\cite{Cho:2015yza}
\bibitem{Cho:2015yza}
  I.~Cho and N.~K.~Singh,
  %``Primordial Power Spectra of EiBI Inflation in Strong Gravity Limit,''
  Phys.\ Rev.\ D {\bf 92}, no. 2, 024038 (2015)
  [arXiv:1506.02213 [gr-qc]].
  %%CITATION = ARXIV:1506.02213;%%


%\cite{Ade:2015tva}
\bibitem{Ade:2015tva}
  P.~A.~R.~Ade {\it et al.} [BICEP2 and Planck Collaborations],
  %``Joint Analysis of BICEP2/$Keck ?Array$ and $Planck$ Data,''
  Phys.\ Rev.\ Lett.\  {\bf 114}, 101301 (2015)
  [arXiv:1502.00612 [astro-ph.CO]].
  %%CITATION = ARXIV:1502.00612;%%


%\cite{Ade:2015lrj}
\bibitem{Ade:2015lrj}
  P.~A.~R.~Ade {\it et al.}  [Planck Collaboration],
  %``Planck 2015 results. XX. Constraints on inflation,''
  arXiv:1502.02114 [astro-ph.CO].
  %%CITATION = ARXIV:1502.02114;%%


\bibitem{Time}
In this article, we use three different time coordinates
as in Refs.~\cite{Cho:2014ija,Cho:2014jta,Cho:2014xaa,Cho:2015yza}.
They are the cosmological time $t$ and the conformal time $\eta$
for the metric, and the conformal time $\tau$ for the auxiliary metric.
At the attractor stage, we have $d\tau = d\eta$ at leading order.
The perturbation fields are in the canonical form in terms of $\tau$.
We begin with the background fields $a$ and $\varphi_0$ in $t$
and the perturbation fields in $\eta$,
and finally study all the quantities in $\tau$.
The beginning of the Universe corresponds to $\tau=0$ ($t\to -\infty$),
and the end of inflation to $\tau=\tau_0 >0$.


%\cite{EscamillaRivera:2012vz}
\bibitem{EscamillaRivera:2012vz}
  C.~Escamilla-Rivera, M.~Banados and P.~G.~Ferreira,
  %``A tensor instability in the Eddington inspired Born-Infeld Theory of Gravity,''
  Phys.\ Rev.\ D {\bf 85}, 087302 (2012)
  [arXiv:1204.1691 [gr-qc]].
  %%CITATION = ARXIV:1204.1691;%%


%\cite{Matsumura:2013aja}
\bibitem{Matsumura:2013aja}
  T.~Matsumura, Y.~Akiba, J.~Borrill, Y.~Chinone, M.~Dobbs, H.~Fuke, A.~Ghribi and M.~Hasegawa {\it et al.},
  %``Mission design of LiteBIRD,''
  Journal of Low Temperature Physics September 2014, Volume 176,
  Issue 5-6, pp 733-740
  [arXiv:1311.2847 [astro-ph.IM]].
  %%CITATION = ARXIV:1311.2847;%%


\bibitem{IP}
Although the background Universe can begin in the past infinity in EiBI inflation,
it was assumed that the initial perturbations were produced
at some moment at the near-MPS stage
in Refs.~\cite{Cho:2014ija,Cho:2014jta,Cho:2014xaa,Cho:2015yza}.
The production mechanism was described in detail in Ref.~\cite{Cho:2015yza}.
The main reasons to consider the specific moment of production are
that the background scalar field may experience large quantum fluctuations
in the course of evolution, so the initial conditions are to be reset at the near-MPS stage,
and that the wavelength scale of the perturbation
is not supposed to be smaller than the Planck scale $l_{p}$.
The latter condition becomes
\be
\lambda_{\rm phys} = \frac{a(\tau_*)}{k} \gtrsim l_p
\qquad\Rightarrow\qquad
\tau_* \gtrsim a^{-1} (kl_p)
\approx \sqrt{-\frac{3\kappa\psi_0}{2a_0^2}}
\left[ \frac{kl_p}{a_0(2\kappa)^{1/3}} \right]^{\sqrt{3/2\kappa m^2}}, \nonumber
\ee
where $\psi_0$ is a negative constant that appears in the near-MPS background solutions.


%\cite{Lagos:2013aua}
\bibitem{Lagos:2013aua}
  M.~Lagos, M.~Banados, P.~G.~Ferreira and S.~Garcia-Saenz,
  %``Noether Identities and Gauge-Fixing the Action for Cosmological Perturbations,''
  Phys.\ Rev.\ D {\bf 89}, 024034 (2014)  [arXiv:1311.3828 [gr-qc]].
  %%CITATION = ARXIV:1311.3828;%%  %5 citations counted in INSPIRE as of 03 Jul 2014

%\cite{Avelino:2012ue}
\bibitem{Avelino:2012ue}
  P.~P.~Avelino and R.~Z.~Ferreira,
  %``Bouncing Eddington-inspired Born-Infeld cosmologies: an alternative to Inflation ?,''
  Phys.\ Rev.\ D {\bf 86}, 041501 (2012)
  [arXiv:1205.6676 [astro-ph.CO]].
  %%CITATION = ARXIV:1205.6676;%%
  %8 citations counted in INSPIRE as of 06 May 2013

%\cite{Yang:2013hsa}
\bibitem{Yang:2013hsa}
  K.~Yang, X.~-L.~Du and Y.~-X.~Liu,
  %``Linear perturbations in Eddington-inspired Born-Infeld gravity,''
  Phys.\ Rev.\ D {\bf 88}, 124037 (2013)  [arXiv:1307.2969 [gr-qc]].
  %%CITATION = ARXIV:1307.2969;%%  %4 citations counted in INSPIRE as of 07 Mar 2014


%\cite{Pani:2011mg}
\bibitem{Pani:2011mg}
  P.~Pani, V.~Cardoso and T.~Delsate,
  %``Compact stars in Eddington inspired gravity,''
  Phys.\ Rev.\ Lett.\  {\bf 107}, 031101 (2011)
  [arXiv:1106.3569 [gr-qc]].
  %%CITATION = ARXIV:1106.3569;%%

%\cite{Pani:2012qb}
\bibitem{Pani:2012qb}
  P.~Pani, T.~Delsate and V.~Cardoso,
  %``Eddington-inspired Born-Infeld gravity.
  %Phenomenology of non-linear gravity-matter coupling,''
  Phys.\ Rev.\ D {\bf 85}, 084020 (2012)
  [arXiv:1201.2814 [gr-qc]].
  %%CITATION = ARXIV:1201.2814;%%

%\cite{Avelino:2012ge}
\bibitem{Avelino:2012ge}
  P.~P.~Avelino,
  %``Eddington-inspired Born-Infeld gravity: astrophysical and cosmological constraints,''
  Phys.\ Rev.\ D {\bf 85}, 104053 (2012)  [arXiv:1201.2544 [astro-ph.CO]].
  %%CITATION = ARXIV:1201.2544;%%

%\cite{Avelino:2012qe}
\bibitem{Avelino:2012qe}
  P.~P.~Avelino,
  %``Eddington-inspired Born-Infeld gravity: nuclear physics constraints and the validity of the continuous fluid approximation,''
  JCAP {\bf 1211}, 022 (2012)  [arXiv:1207.4730 [astro-ph.CO]].
  %%CITATION = ARXIV:1207.4730;%%  %26 citations counted in INSPIRE as of 10 sept. 2015

%\cite{Pani:2012qd}
\bibitem{Pani:2012qd}
  P.~Pani and T.~P.~Sotiriou,
  %``Surface singularities in Eddington-inspired Born-Infeld gravity,''
  Phys.\ Rev.\ Lett.\  {\bf 109}, 251102 (2012)  [arXiv:1209.2972 [gr-qc]].
  %%CITATION = ARXIV:1209.2972;%%

%\cite{Sotiriou:2008rp}
\bibitem{Sotiriou:2008rp}
  T.~P.~Sotiriou and V.~Faraoni,
  %``f(R) Theories Of Gravity,''
  Rev.\ Mod.\ Phys.\  {\bf 82}, 451 (2010)  [arXiv:0805.1726 [gr-qc]].
  %%CITATION = ARXIV:0805.1726;%%  %1352 citations counted in INSPIRE as of 10 sept. 2015

%\cite{Kim:2013nna}
\bibitem{Kim:2013nna}
  H.~C.~Kim,
  %``Physics at the surface of a star in Eddington-inspired Born-Infeld gravity,''
  Phys.\ Rev.\ D {\bf 89}, no. 6, 064001 (2014)  [arXiv:1312.0705 [gr-qc]].
  %%CITATION = ARXIV:1312.0705;%%  %15 citations counted in INSPIRE as of 10 sept. 2015

\end{thebibliography}
\end{document}